\documentclass[conference,letterpaper]{IEEEtran}

\usepackage[utf8]{inputenc} 
\usepackage[T1]{fontenc}
\usepackage{url}
\usepackage{ifthen}
\usepackage{cite}
\usepackage[cmex10]{amsmath} % Use the [cmex10] option to ensure complicance
\usepackage{float}
\usepackage{graphicx}
\usepackage{caption}
\usepackage{comment}
\usepackage{xcolor}

\begin{document}
\title{On the Sensitivity to Errors in Homomorphic Computing: Single Transient Bit-flip Client-side Error Characterization}

\author{
    \IEEEauthorblockN{Matías Mazzanti\textsuperscript{1}, Vattana Chan\textsuperscript{3}, Karthik Swaminathan\textsuperscript{2}, Augusto Vega\textsuperscript{2}, Esteban Mocskos\textsuperscript{1}, Radha Venkatagiri\textsuperscript{3}\\}
    \IEEEauthorblockA{
        \textit{\textsuperscript{1}University of Buenos Aires, \textsuperscript{2}IBM T. J. Watson Research Center, \textsuperscript{3} Georgetown University}}
}
\maketitle

\begin{abstract}
Homomorphic Encryption (HE) enables computation on encrypted data without decryption and is a key primitive for privacy-preserving computation in sensitive domains such as healthcare, finance, and government. Its security relies on noise injection, which introduces intrinsic error sensitivity and raises concerns about the fault tolerance of HE systems, as hardware- and software-induced faults can evade traditional detection mechanisms and lead to silent data corruption.

In this work, we analyze the sensitivity of HE to bit-level faults, focusing on the CKKS (Cheon--Kim--Kim--Song) scheme widely used for approximate arithmetic in AI and machine learning workloads. We identify homomorphic multiplication as the most error-sensitive operation in practical HE pipelines and characterize how faults propagate and amplify through it, exposing a critical robustness vulnerability and motivating the need for more resilient HE deployments.
\end{abstract}

\section{Introduction and Background}
\label{sec:introduction}

\begin{figure}[!ht]
  \centering
    \includegraphics[width=0.95\columnwidth]{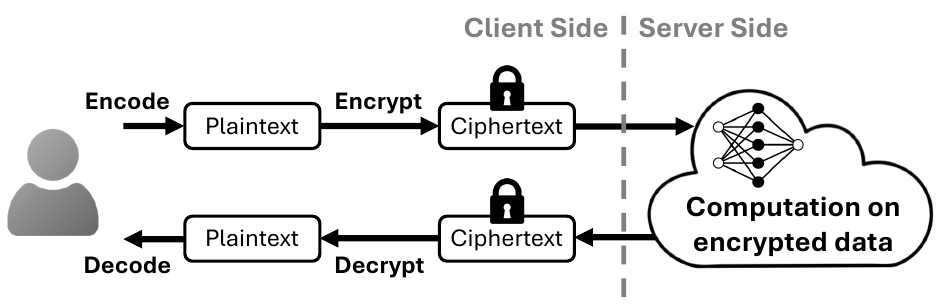}
	\caption{Homomorphic encryption used in a typical business application: users send their data encrypted to a third-party service provider, where data is processed in its encrypted form.}
	\label{fig:he_overview}
\end{figure}

Cloud computing has profoundly changed the way businesses and individuals use, process, and manage their data \cite{Konstantinos2015}. Despite the benefits of this approach, its adoption exacerbates some existing problems and generates new ones related to the security and privacy of user data~\cite{Sen2015}. In particular, delivering data to a third-party to be processed opens up different possibilities of compromising them, both by improper access or by a decision of the provider to give access to the data without authorization.
One of the solutions to this problem resides in the use of data encryption schemes, which allow the data to only be interpreted by the holder of the key that allows decryption~\cite{Williams1980,Elgamal1986}.
However, most cryptographic schemes do not allow computation directly on encrypted data; instead, the data must be decrypted before processing.
In this way, an external cloud computing provider can access the unencrypted data and; therefore, be able to compromise them. 
\textit{Homomorphic Encryption} (HE) solves this challenge by allowing \textbf{computation on encrypted data}~\cite{Yi2014}. 
Although HE schemes have high computational and memory requirements, which have limited so far their widespread adoption, the great interest in developing new schemes and the recent development of aggressive optimizations (at both algorithm and hardware levels) has opened the door to its use in real-world applications. Figure~\ref{fig:he_overview} presents an overview of a typical HE setting that involves a client that makes use of third-party cloud services.

Multiple HE schemes exist today, with different characteristics and supported capabilities. In this work, we focus on CKKS (Cheon-Kim-Kim-Song)~\cite{Cheon2017,Cheon2019}, a popular HE scheme for AI and machine learning applications due to its floating-point arithmetic support. Fundamentally, schemes like CKKS base their hardness on a simple idea: add small errors (``noise'') to data to make relatively simple problems computationally intractable, an approach known as \textit{Learning With Errors} (LWE)~\cite{10.1145/1060590.1060603}. In other words, HE operations (such as additions and multiplications \textit{etc.}) take place within a ``noisy'' domain.

In this paper, we present that 1). error resilience of FHE on client-side can be categorized into two: Addition pattern and Multiplication pattern across Vanilla and RNS, 2). Multiplication pattern dominates addition pattern when a multiplication is performed on the server and 3). the two resilience patterns observed are parametric and data independent.

\subsection{Silent Data Corruption}
% \footnote{The FHE library used in this work (OpenFHE) detects such data alterations and terminates execution with an assertion error.}
Errors occurring at different stages of CKKS execution—encoding, encryption, decryption, and decoding can lead to three outcomes. Either the HE execution fails and the error is explicitly \textit{detected} at decoding, the execution completes successfully while the error propagates across stages and manifests as \textit{silent data corruption (SDC)} \cite{dixit2021silentdatacorruptionsscale}, or the error may be \textit{masked} and have no observable effect on the output.

\begin{figure}[!ht]
  \centering
    \includegraphics[width=0.8\columnwidth]{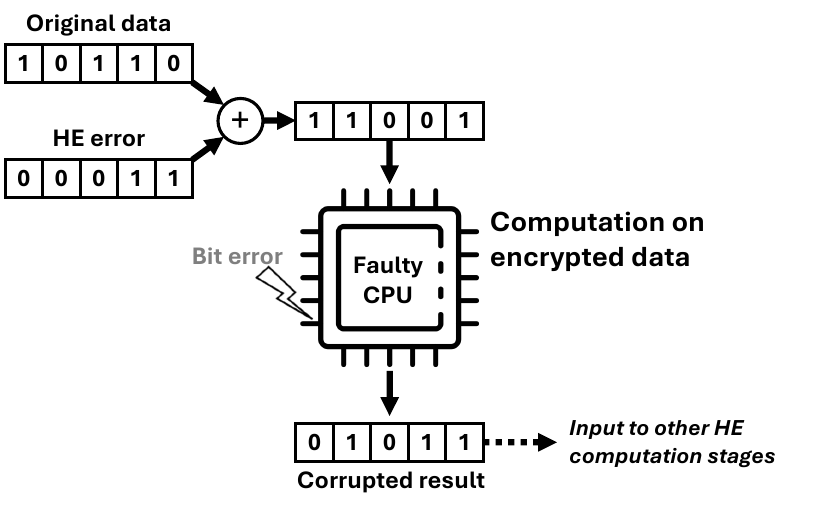}
	\caption{Illustrative scenario of a SDC case induced by a faulty CPU. The corrupted result incorporates both the HE error and the faulty hardware error.}
	\label{fig:he_error_example}
\end{figure}

The severity of SDC in HE applications stems from the error-centric nature of HE schemes themselves. In the encrypted domain, data is inherently noisy by construction; consequently, additional errors induced by faulty hardware or software can become indistinguishable from the intrinsic HE noise, making detection extremely challenging. Figure~\ref{fig:he_error_example} illustrates this effect, where plaintext data is encrypted by injecting random HE noise\footnote{In practice, encryption involves additional steps not depicted in the figure.}.

\begin{comment}

\subsection{Silent Data Corruption}
Errors across CKKS stages (encoding, encryption, decryption, and decoding) can result in two scenarios: HE operation ``breaks''\footnote{The FHE library used in this work (OpenFHE) detects the data alteration and finishes its execution with an assertion error.} and the error is detected, or HE operation does not break and the error propagates across stages and ends up on \textit{silent data corruption} (SDC). The latter is the dangerous case and, as it has been widely studied and reported, hardware- and software-induced SDC happens, even in today's cutting edge systems and large-scale datacenters~\cite{dixit2021silentdatacorruptionsscale}. The nasty aspect about hardware and software errors in HE applications emerges from the very same error-centric intrinsic operation of HE schemes. Once in the HE domain, data becomes noisy by construction and, if additional error occurs due to faulty hardware or software, such error camouflages within the HE error and becomes very hard to detect. Figure~\ref{fig:he_error_example} presents a cartoonish illustration of this idea, where original data (plaintext) is ``encrypted'' by adding random HE error (noise)\footnote{In practice, encryption involves additional steps not shown in the figure.}.
    
\end{comment}

\section{Error Resilience Analysis}
\label{sec:error_resilience_analysis}

This section presents preliminary error sensitivity results for CKKS. As illustrated in Figure~\ref{fig:he_overview}, the \textit{encoding} stage transforms user's input data into a \textit{plaintext (ptx)}, a polynomial of degree $N$ that we will call $p(X)$. This plaintext is then encrypted into a \textit{ciphertext}, a pair of polynomials of degree $N$ each, that we will call $c = (c_0(X), c_1(X))$. We adopt a single-bit error model. As such, each bit of every polynomial coefficient (in both the plaintext and the ciphertext) is flipped in sequence for the set of single-bit-flip fault injection experiments.
After each bit error injection, we execute the entire HE pipeline, and compare the recovered data after decoding (last stage) against the original data using the Maximum Relative Error Percentage (MREP). %This methodology is depicted in Figure~\ref{fig:methodology}, where two input elements (original message) are encoded into a 4-element polynomial (plaintext) and encrypted into two 4-element polynomials (ciphertext). 
%A coefficient bit is flipped at a time before decryption and decoding. 
%The recovered message is compared against the original one using the Maximium Relative Error (MRE). 
We executed all experiments on an Intel i7-11700 CPU with 32 GB RAM and Arch Linux 257.5-1.

\textbf{FHE optimizations and parameters.} We use a custom C++ implementation of CKKS (C-CKKS) \footnote{C-CKKS is implemented based on OpenFHE \cite{AlBadawi2022}, HEaaN \cite{Cheon2017}, SEAL \cite{MicrosoftSeal} and PyFHE \cite{pyfhe}.}that supports native RNS, NTT, combined RNS+NTT arithmetic, and 64-bit coefficient representations. 
We refer to the configuration where both RNS and NTT optimizations are disabled as \emph{Vanilla CKKS}. 
All optimizations can be selectively enabled or disabled, enabling a controlled and fine-grained evaluation. 
This design allows systematic variation of critical FHE parameters such as the modulus \textit{Q}, scaling factor \textit{$\Delta$}, ring dimension  \textit{N}, and slot/gap configurations—providing insight into CKKS behavior under error-driven conditions. Fault injection experiments are conducted using the open-source LLTFI \cite{LLTFI} framework.

%\textbf{FHE optimizations and parameters} Using our own CKKS implementation (C-CKKS) in C++, which includes native RNS, NTT, RNS+NTT and 64-bit coefficient representation support with the option to turn off these optimizations for a more detailed and robust study especially with the ability to vary critical FHE parameters such as modulus \textit{Q}, scaling factor \textit{$\Delta$}, ring dimension \textit{N}, including slots/gaps, generating us interesting insights how FHE behaves under error-driven environment. In combination with an open source error-injection tool called LLTFI.

\textbf{FHE pipeline configurations.} We experimented FHE under various combinations of basic FHE computational operations such as No Computation (NoComp), Addition (Add), Multiplication (Mult), Rotation (Rot), combinations of basic computational operations (i.e Add + Rot, Add + Mult, Rot + Mult \textit{etc.}) these basic computational blocks followed by boostrapping (Add + Mult + Boot).

% \begin{figure}[!ht]
%   \centering
%     \includegraphics[width=1\columnwidth]{figures/methodology.pdf}
% 	\caption{Error injection methodology.}
% 	\label{fig:methodology}
% \end{figure}

\subsection{FHE Error Resilience}
Error resilience can be categorized into two: 1). Addition pattern and 2). Multiplication pattern.
% \subsection{Vanilla CKKS}

\textbf{Addition Pattern.}
We observe that, in Vanilla CKKS, bit-flips affecting values below the scaling factor (\(\Delta\)) or exceeding the modulus (\(Q\)) resilience lead to masked plaintext and both ciphertext components, \(c_0\) and \(c_1\). Bit-flips occurring on gap coefficients or on the \(\frac{N}{2}\)-th coefficient consistently produce fully masked effects in the plaintext and in \(c_0\), but not in \(c_1\). This behavior arises because gaps are present in the plaintext and are therefore encoded into \(c_0\), whereas \(c_1\) corresponds to a sampled secret-key-dependent component. A closely related pattern is observed in RNS CKKS; however, individual coefficients do not exhibit scaling-factor (\(\Delta\)) or modulus (\(Q\)) resilience, as each coefficient is represented using multiple limbs.

%We observe that bit-flips below scaling factor (\textit{$\Delta$}) and above modulus (\textit{Q}) resilience in Vanilla CKKS lead to masked results for plaintext, $c_0$ and $c_1$. Bit-flips occur on gap and $\frac{N}{2}^{th}$ coefficients will always lead to fully masked results only in plaintext, $c_0$, but not $c_1$. This is due to the fact that gaps exist in plaintext, and encrypted into $c_0$ while $c_1$ is a sampled-key. A closely similar pattern is also seen in RNS CKKS with the exception that coefficients do not have scaling factor (\textit{$\Delta$}) and above modulus (\textit{Q}) resilience because each coefficient is represented by limbs.

\begin{figure}[H]
    \centering
    \includegraphics[width=\columnwidth,keepaspectratio]{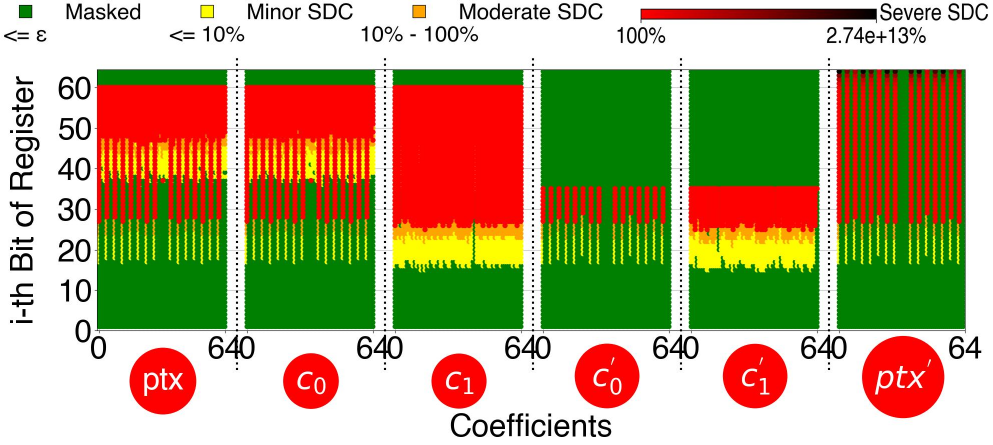}
    \caption{Error characterization of $CKKS + Mult_1$ with Gaps = 2, $\log\text{Q}$ = 60, $\log\Delta$ = 25, N = 64, $\varepsilon = 0.1$. \textit{ptx'}, \textit{$c_0^{'}$}, and \textit{$c_1^{'}$} denote computed ciphertext-ciphertext \textit{$c_0$}, \textit{$c_1$}, and decrypted \textit{ptx} respectively.}
    \label{fig:gap-config}
\end{figure}

\textbf{Multiplication pattern.} 
An error-resilience profile closely resembling that of addition is observed in both Vanilla CKKS and RNS CKKS. However, in Vanilla CKKS, gap and \(\frac{N}{2}^{\text{th}}\) coefficients remain resilient only up to a certain error threshold (see Fig.~\ref{fig:gap-config}), whereas in RNS CKKS these coefficients exhibit no resilience and are completely corrupted. 

In contrast, CKKS configurations employing NTT and RNS+NTT arithmetic are fully susceptible to bit-flips, as any single bit-flip ultimately results in catastrophic decryption and decoding failures. Consequently, no distinct addition or multiplication error patterns can be identified under these configurations.

%\textbf{Multiplication Pattern.} An error resilience profile almost identical to that of addition is observed Vanilla CKKS and RNS CKKS; however, the gap and $\frac{N}{2}^{\text{th}}$ coefficients remain resilient only up to a certain threshold in Vanilla CKKS (see Fig.~\ref{fig:gap-config}) while the gaps and $\frac{N}{2}^{\text{th}}$ coefficients are completely destroyed in RNS (no resilience at all for all coefficients).

 %FHE in NTT and RNS+NTT are fully susceptible to bit-flips as every single bit flip leads to eventual catastrophic results in decryption and decoding; therefore, we do not observe addition and multiplication patterns.

\subsection{Multiplication Pattern Domination}

Our observations indicate that whenever a multiplication is performed on the server, the overall resilience profile consistently follows this pattern.

% \subsection{RNS, NTT and RNS+NTT Optimization}
% \subsubsection{RNS} No resilience in respect to scaling factor (\textit{$\Delta$}), and modulus (\textit{Q}) are observed on all cofficients. Althought fascinatingly the gaps and $\frac{N}{2}^{th}$ resilience are still preserved even in RNS domain, these gaps and $\frac{N}{2}^{th}$ are completely destroyed once a multiplication is performed on the server.
\subsection{Parameter and Data Independence}
Interestingly, the observed resilience heuristics are intrinsic to the scheme itself, rendering them independent of both parameters and data.

\section{Conclusion}
In this particular work, we bring a formal awareness of how errors that occur in the underlying hardware may corrupt the effectiveness of FHE under different parameters and configurations. First of all, error resilience of FHE on client-side can be formally categorized into Addition pattern and Multiplication pattern. Secondly, whenever a multiplication is performed on the server, a multiplication pattern dominates the pattern of addition. Finally, these resilience profiles categorized are data and parameter independent.

\bibliographystyle{IEEEtran}
\bibliography{refs}
\end{document}